\documentclass[11pt]{article}

\def\mytitle#1{\setcounter{equation}{0}
\setcounter{footnote}{0}
\begin{flushleft}\Large\textbf{#1}\end{flushleft}
\vspace{0.25cm}}
\def\myname#1{\leftline{{\large #1}}\vspace{-0.13cm}}
\def\myplace#1#2{\small\begin{flushleft}\textit{#1}\\
\texttt{#2}\end{flushleft}}

\def\myclassification#1{\small\noindent
Pacs no :
       #1\vspace{0.5cm}}
\usepackage{graphicx}
\begin{document}

\mytitle{Universal Thermodynamics with state finder parameters: A
general prescription}

 \myname{Nairwita
Mazumder\footnote{nairwita15@gmail.com}} \vskip0.2cm
\myname{Ritabrata Biswas\footnote{biswas.ritabrata@gmail.com}}
\vskip0.2cm \myname{Subenoy
Chakraborty\footnote{schakraborty@math.jdvu.ac.in}} \vskip0.2cm
\myplace{Department of Mathematics, Jadavpur University,
Kolkata-700 032, India.} { }

\begin{abstract}
The work deals with universal thermodynamics for flat FRW model,
considering interacting $n-$fluid system. By introducing the state
finder parameters, the feasible regions in the $\{r,s\}-$plane has
been examined for the validity of the Generalized Second Law of
Thermodynamics. A general prescription for the evolution of the
horizon has been derived for general non-static spherically
symmetric space time and then it is applied to FRW universe. A
general restriction on the matter has been found for the validity
of GSLT.
\end{abstract}
\myclassification{98.80.Cq,95.35.+d,98.80.-k}
\section{Introduction}
In 1974, Hawking \cite{Hawking1,Hartle1} discovered that a black
hole(BH) behaves like a black body, emmitting thermal radiations
with a temperature proportional to its surface gravity at the
horizon and the related entropy is proportional to its horizon
area \cite{Bekenstein1,Bardeen1}. This entropy, temperature and
the mass of the BH are related by the first law of BH
thermodynamics $(dM=T dS)$\cite{Bardeen1,Gibbons1}. Due to the
above thermodynamics-geometry relationship i.e.
$$(temperature,~Entropy)\approx (Surface~gravity,~Area),$$
There had been speculations in the literature that thermodynamic
laws are in some how related to the Einstein equations. The real
picture came in 1995 when Jacobson \cite{Jacobson1} was able to
formulate Einstein equations from the thermodynamical laws.
Subsequently, Padmanabhan \cite{Padmanabhan1,Paranjape1} showed
the equivalence in the otherway. Most discussions on BH
thermodynamics have been related to static BH. Subsequently,
Hayward \cite{Hayward1,Hayward2} have initiated the study of
non-static BH (dynamic BH) by introducing the concept of trapping
horizon. For spherically symmetric space-times he was able to
write down the Einstein field equations in a form, termed as
'Unified first law' and first law of thermodynamics for a
dynamical BH can be obtained by projecting this unified first law
along a trapping horizon. Subsequently, a lot of works has been
going on related to universal thermodynamics. These studies
\cite{Wang1,Cai1,Cai2,Akbar1,Karami1,Mazumder1,Mazumder2,Mazumder3,Mazumder4,Mazumder5}
are mostly concentrated to homogeneous and isotropic FRW model of
the universe bounded by apparent or event horizon and conditions
for the validity of generalized second law of thermodynamics(GSLT)
have been
evaluated.\\

In the present work, we formulate a general prescription of
universal thermodynamics for FRW model starting from an
interacting $n-$fluid system. by introducing, the state finder
parameters, we derive the feasible region in the ${r,s}-$plane for
the validity of GSLT. also formulating the evolution of the
horizon, a general restriction on the matter has been derived and
it is applied to the FRW model.\\

\section{Universal Thermodynamics: A General Prescription}

We consider a homogeneous, isotropic, flat, FRW model of the
universe. Suppose the universe is filled up with interaction
n-fluids having energy densities and thermodynamic pressures are
$(\rho_i,p_i),i=1,2,...n$. So the conservation equations are
\begin{equation}\label{1}
{\dot{\rho}}_i+3H(\rho_i+p_i)=Q_i,~i=1,2,...1
\end{equation}

where $Q_i$ stands for the interaction term related to $i-$th
fluid it may be positive or negative so that
$\sum_{i=1}^{n}Q_i=0$. Now combining the $n-$equations in
$(\ref{1})$ we have
\begin{equation}\label{2}
\dot{\rho}+3H(\rho+p)=0
\end{equation}

where $\rho=\sum_{i=1}^{n}{\rho}_i,$ is the total energy density
of the $n-$fluids and $p=\sum_{i=1}^{n}{p}_i,$ is the total
thermodynamic pressure of the $n-$fluid. Thus equation (\ref{2})
can be considered as the energy conservation equation of the
resulting single fluid having energy
density $\rho$ and thermodynamic pressure $p$.\\

The Einstein field equations for the above interacting $n-$fluid
system are given by

\begin{equation}\label{3}
3H^2=\rho
\end{equation}
and
\begin{equation}\label{4}
2\dot{H}=-(\rho+p)
\end{equation}

To study the thermodynamics of the universe as a thermodynamical
system we start with the amount of energy crossing the horizon
$(R_h)$ during an infinitesimal interval of the time $dt$ as
\begin{equation}\label{5}
-dE=4 \pi R_h^2 T_{\mu \nu}l^{\mu}l^{\nu}dt= 4 \pi R_h^3
H(\rho+p)dt
\end{equation}

Then from the clausius relation

\begin{equation}\label{6}
-dE=dQ=T_hdS_h \end{equation}

We have,
\begin{equation}\label{7}
\frac{dS_h}{dt}=\frac{4 \pi R_h^3H}{T_h}(\rho+p)
\end{equation}

where $dQ$ is the heat flux crossing the horizon during the time
interval $dt$, $T_h$ and $S_h$ are respectively the temperature
and entropy at the horizon.\\

Now the entropy $(S_I)$ of the matter of the universe inside the
horizon can be related to its energy and pressure in the horizon
by Gibb's equation

\begin{equation}\label{8}
TdS_I=dE_I+pdV
\end{equation}

where $S_I=\sum_{i=1}^{n}S_i$ is the sum of the entropies of
different matter components, $V=\frac{4}3 \pi R_h^3$ is the volume
of the universe enclosed by the horizon and $E_I=\rho.V$ is the
total energy of the matter bounded by the horizon. Then using the
conservation equation (\ref{2}) and after a little bit of
calculations we obtain the variation of the inside matter entropy
as
\begin{equation}\label{9}
\frac{dS_I}{dt}=\frac{1}{T_h}\left[-4 \pi R_h^3 H (\rho+p)+4 \pi
R_h^2(\rho+p){\dot{R}}_h\right]
\end{equation}
where for equilibrium thermodynamics we chose the temperature of
the inside matter same as that of the horizon. Thus combining
(\ref{7}) and (\ref{9}) the time variation of total entropy
(matter and horizon) is given by

\begin{equation}\label{10}
\frac{dS_{tot}}{dt}=\frac{4 \pi R_h^2}{T_h}{\dot{R}}_h (\rho+p)
\end{equation}

Thus generalized second law of thermodynamics (GSLT) (i.e.
$\frac{dS_{tot}}{dt}\geq0$) will always be true for ay one of the
following two possibilities:\\

(a)The equivalent single fluid for the interacting $n-$fluid
 system is in quintessence era and horizon is non-decreasing with
 the evolution.\\

 (b) The equivalent single fluid system is in phantom era and
 horizon is contracting with the evolution.\\

 It should be noted that the validity of the GSLT does not depend
 on any individual fluid nor on the interaction among the fluids.\\

 Now we shall express the universal thermodynamics in terms of
 state finder parameters $\left\{ r,s\right\}$, proposed by Sahni
 et al \cite{Sahni1,Debnath1}. According to them, statefinder
 parameters are defined as

 \begin{equation}\label{11}
 r=\frac{1}{aH^3}\frac{d^3a}{dt^3}~and~s=\frac{(r-1)}{3(q-\frac{1}2)}
 \end{equation}

 where $q=-\frac{\ddot{a}a}{{\dot{a}}^2}$ is the deceleration
 parameter.\\

 The parameters $`r'$ and $`s'$ are dimensionless and $`r'$ forms
 the next step i the hierarchy of geometrical cosmological
 parameters after $H$ and $q$ \cite{Sahni1}. It is speculated
 that these parameters together with coming SNAP observations may
 discriminate between different dark energy models. Further,
 trajectories in the $\left\{r,s\right\}$ plane corresponding to
 different cosmological models demonstrate qualitatively different
 behaviour.\\

 Now using these parameters $q,r,~and~s,$ the rate of change of
 the Hubble parameter can be written as

 \begin{equation}\label{12}
 \dot{H}=-(1+q)H^2=-\frac{3H^2}2\left[1+\frac{2(r-1)}{9s}\right]
\end{equation}

So from the Einstein field equation (\ref{4}) we have

\begin{equation}\label{13}
\rho+p=3H^2\left(1+\frac{2(r-1)}{9s}\right)
\end{equation}
Thus variation of total entropy can be expressed in terms of
$\left\{r,s\right\}$ parameters as

\begin{equation}\label{14}
\frac{dS_{tot}}{dt}=\frac{12 \pi R_h^2
H^2}{T_h}\left\{1+\frac{2(r-1)}{9s}\right\}{\dot{R}}_h
\end{equation}
\begin{figure}
\includegraphics[height=2in, width=2in]{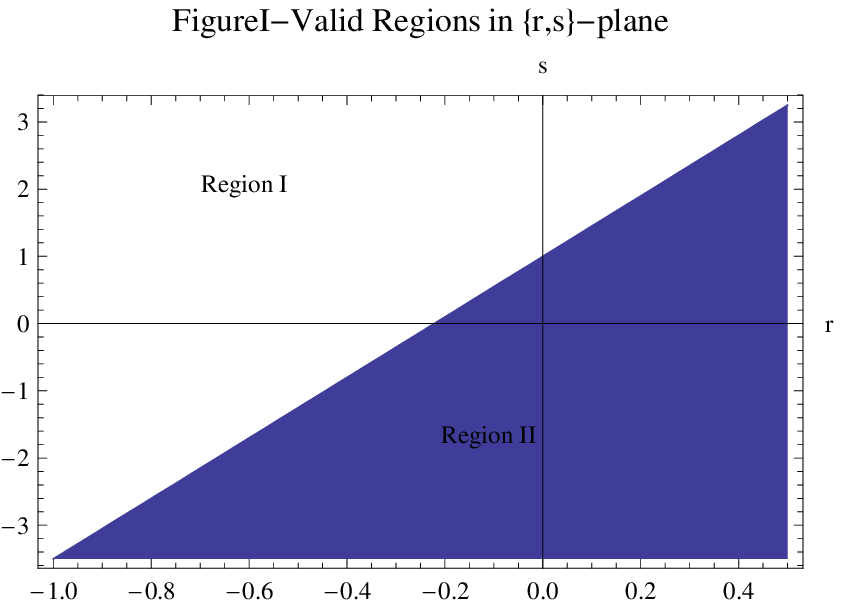}~~~~
\vspace{1mm} \vspace{6mm}
\end{figure}
If we now draw the straight line $2(r-1)+9s=0$ in the
$\{r,s\}-$plane then region I (see Fig1) corresponds to the valid
region for GSLT to be satisfied provided horizon radius grows with
the evolution of the universe otherwise region II will be the
possible region for validity of GSLT. Therefore, we can classify
regions in the $\{r,s\}-$plane where universe can be
considered as a thermodynamical system.\\

Further, for the commonly used horizon in the literature,namely
the apparent horizon given by\\

\begin{equation}\label{15}
R_A=\frac{1}H
\end{equation}

we have
\begin{equation}\label{16}
{\dot{R}}_A=-\frac{1}{H^2}\dot{H}=\frac{3}2\left\{1+\frac{2(r-1)}{9s}\right\}
\end{equation}
So
\begin{equation}\label{17}
\frac{dS_{tot}}{dt}=\frac{18 \pi R_h^2
H^2}{T_h}{\left\{1+\frac{2(r-1)}{9s}\right\}}^2\geq0
\end{equation}

Thus we have the known result that GSLT is always satisfied at the
apparent horizon.\\

On the other hand, the event horizon (which is defined for
accelerating model) is defined as

\begin{equation}\label{18}
R_E=a\int_t^{\infty}\frac{dt}a
\end{equation}
and we have

$${\dot{R}}_E=H(R_E-R_A)$$

and it can not be expressed in terms of
$\left\{r,s\right\}-$parameters. So in this case $R_E>R_A$ and
region I or region II
 is the possible solution for the validity of GSLT. However, if we
 use the Ricci's length scale namely

\begin{equation}\label{19}
 R_L={2H^2+\dot{H}}^{\frac{-1}2}
\end{equation}

as the boundary of the universe, then
\begin{equation}\label{20}
{\dot{R}}_L=\frac{1}2R_L^3H^3(q+2-r)=\frac{1}2R_L^3H^3
\left\{\frac{3}2-(r-1)(1-\frac{1}{3s})\right\}
\end{equation}
\begin{center}
\begin{figure}
\includegraphics[height=2.5in, width=2.2in]{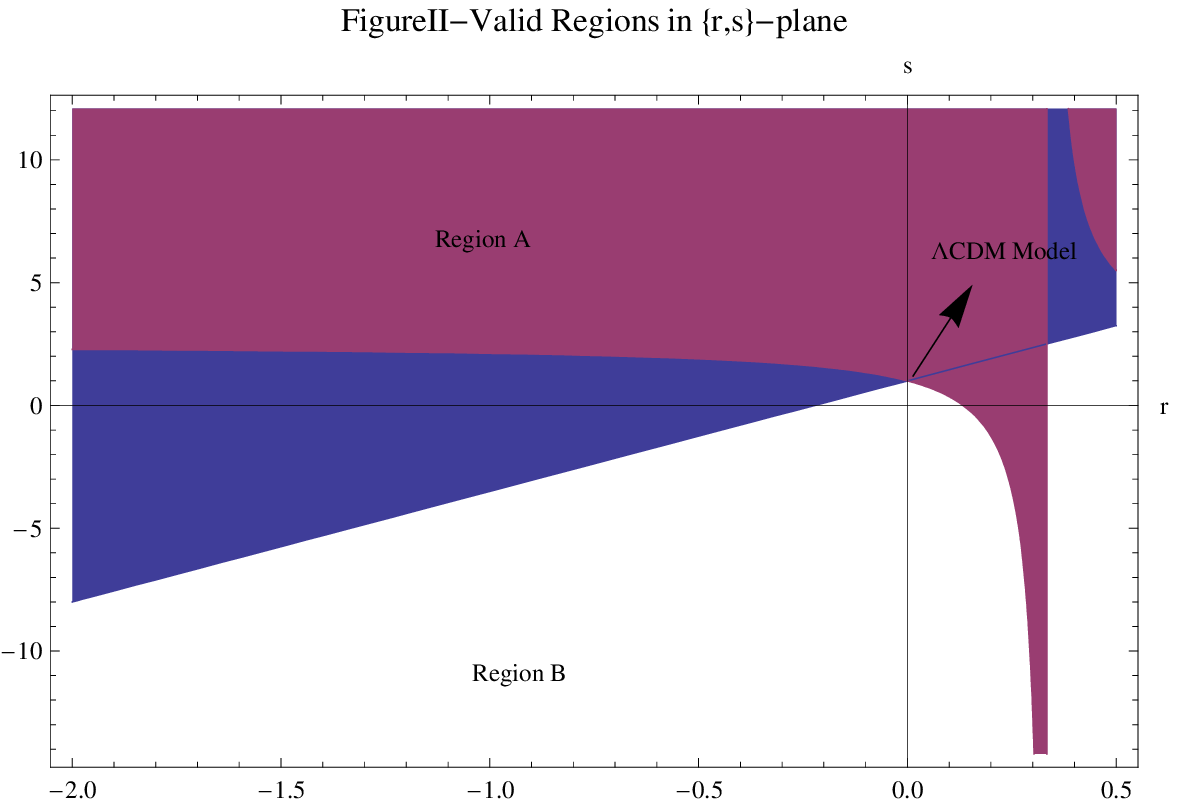}~~~~
\vspace{1mm} \vspace{6mm}
\end{figure}
\end{center}
The argument behind $R_L$ as the IR cut off in the holographic
bound \cite{Lepe1,Duran1} is that it corresponds to the size of
the maximal perturbation, leading to the formation of a black
hole. Thus in this case the possible part of the the
$\{r,s\}-$plane for
the validity of GSLT is classified as\\

$$Region~A:~\frac{3}2-(r-1)\left(1-\frac{1}{3s}\right)>0~\mbox{and}~1+\frac{2(r-1)}{9s}>0$$
$$or$$
$$Region~B:~\frac{3}2-(r-1)\left(1-\frac{1}{3s}\right)<0~\mbox{and}~1+\frac{2(r-1)}{9s}<0$$
In figure 2 we have identified regions A and B which are the valid
region for the universal thermodynamics.\\

\section{Evolution of the horizon in general spherically symmetric black hole}

In this section, we study the growth or contraction of the horizon
of a general non-static spherically symmetric black hole due to
exchange of matter with its environment and then we extend this
idea to the universal thermodynamics. Any spherically symmetric
gravitational field can be written as

\begin{equation} \label{21}
dS^2=-f(r,t)dt^2+\frac{1}{g(r,t)}dr^2+r^2d{\Omega}_2^2
\end{equation}

The horizon is located at $r=r_h$ such that $f(r_h,t)=0=g(r_h,t)$.
To have a smooth system across the horizon we make the following
Painleve-type coordinate transformation:

\begin{equation}\label{22}
dt \rightarrow dt-\sqrt{\frac{1-g}{fg}}
\end{equation}

so that metric (\ref{21}) becomes
\begin{equation}\label{23}
dS^2=-f dt^2+2\sqrt{\frac{f(1-g)}{g}}dt dr+dr^2+r^2d{\Omega}_2^2
\end{equation}

The incoming and out going radial null geodesics are defined by
$dS^2=0=d \theta=d \phi$ and so from equation (\ref{23})
\begin{equation}\label{24}
\frac{dr}{dt}=-\sqrt{\frac{f}{g}}\left[\sqrt{1-g}\pm 1\right]
\end{equation}
Now we define the evolving horizon by the condition

\begin{equation}
\sqrt{1-g}=1
\end{equation}

Thus if $\sqrt{1-g}>1$ then out going light rays are dragged
backwards towards origin. As a result we can define a function
$r_h(t)$ such that $g(r_h^{(t)},t)=0$ and is termed as evolving
horizon. We can identify the mass function as

\begin{equation}\label{26}
2m(r,t)=r[1-g(r,t)]
\end{equation}
and hence the horizon

\begin{equation}\label{27}
2m(r_h(t),t)=r_h(t)
\end{equation}

Further due to spherical symmetry, the Hawking-Isreal quasi local
mass function \cite{Nielsen1} namely,

\begin{equation}\label{28}
m_{HI}(r,t)=\frac{r}2\left[1-g^{\mu
\nu}{\nabla}_{\mu}r{\nabla}_{\nu}r\right]
\end{equation}

is identified with $m(r,t)$ defined in (\ref{26}). So we can
identify the function $`m'$ as the 'mass inside radius r at time
t'.\\

Now taking total time derivative of both sides of equation
(\ref{27}) we have

\begin{equation}\label{29}
\dot{m}(r_h(t),t)=\frac{\left[1-2m'(r_h(t),t)\right]}{16 \pi
r_h(t)}\frac{dA_h(t)}{dt}
\end{equation}

where $A_h=4 \pi r_h^2$ is the surface area of the horizon and
here dot and dash stands for partial derivatives with respect to
$`t'$ and $`r'$ respectively.\\

Now comparing with the first law of black hole mechanics namely,

\begin{equation}\label{30}
dm=\frac{k}{8 \pi} dA,
\end{equation}

the surface gravity on the horizon has the expression

\begin{equation}\label{31}
k_h=\frac{\{1-2m'(r_h(t),t)\}}{2r_h(t)}
\end{equation}

It is to be  noted that the surface gravity $k$ can also be
obtained from the evolution equation of outward radial null vector

\begin{equation}\label{32}
l^{\mu}\nabla_{\mu}l^{\nu}=kl^{\nu}
\end{equation}

where
\begin{equation}\label{33}
l^{\mu}=\left\{\sqrt{\frac{g}f},\sqrt{1-g},0,0\right\}
\end{equation}

is the out going radial null vector.\\
As the metric coefficient $g(r,t)$ is positive definite for
$r>r_h(t)$ so from (\ref{26}) $2m(r,t)<r$ for $r>r_h(t)$ or
equivalently
\begin{equation}\label{34}
1-2m'(r_h(t),t)\geq 0~i.e.~r_hg'(r_h,t)\geq 0
\end{equation}

Hence the surface gravity on the horizon is positive definite.
Further, using Maple(or similar software) one can calculate (for
details see ref \cite{........}) for any $r$

\begin{equation}\label{35}
G_{\mu \nu}l^{\mu}l^{\nu}= \frac{2}{r^2}\sqrt{\frac{g}f}
\frac{\dot{m}(r,t)}{\sqrt{1-g}}\frac{2}{r}\sqrt{\frac{g}f}
\frac{\partial}{\partial r}\left\{\sqrt{\frac{f}g}\right\}
{(1-\sqrt{1-g})}^2
\end{equation}

So on the horizon,

\begin{equation}\label{36}
\dot{m}(r_h(t),t)=\frac{r_h^2}{2}\sqrt{\frac{f'(r_h,t)}{g'(r_h,t)}}(8
\pi T_{\mu \nu}l^{\mu}l^{\nu}),
\end{equation}

where in deriving the last equation we have used Einstein
equations $G_{\mu \nu}=8 \pi T_{\mu \nu}$. So from equation
(\ref{29}) we obtain
\begin{equation}\label{37}
\frac{dr_h}{dt}=8 \pi r_h T_{\mu \nu}l^{\mu}l^{\nu}
\sqrt{\frac{f'(r_h(t),t)}{{g'(r_h(t),t)}^3}}
\end{equation}

Further using (\ref{27}) the total time derivative of the mass
function at the horizon gives
\begin{equation}\label{38}
\frac{dm(r_h(t),t)}{dt}=4 \pi r_h T_{\mu
\nu}l^{\mu}l^{\nu}\sqrt{\frac{f'}{{g'}^3}}
\end{equation}

This gives the total mass change i.e. contributions both from mass
flux across the horizon and from the motion of the horizon itself.
Thus due to inequality (\ref{34}) we may conclude that as long as
null energy condition (NEC) is satisfied both the horizon as well
as the mass bounded by the horizon can never decrease.\\

As a particular case if we consider the flat FRW model of the
universe we can write the metric similar to (\ref{23}) as

\begin{equation}\label{39}
dS^2=-(1-H^2R^2)dt^2-2HRdRdt+dR^2+R^2d\Omega_2^2
\end{equation}

So the horizon is located at $R=R_h=\frac{1}H=R_A$, the apparent
horizon.

The mass term is defined as
\begin{equation}\label{40}
m(R,t)=\frac{1}2H^2R^3
\end{equation}
and hence the mass bounded by the horizon is
\begin{equation}\label{41}
m(R_h(t),t)=\frac{R_h(t)}2,
\end{equation}
 which is the usual Mishner-Sharp mass at the apparent horizon of
 the FRW model. Hence from (\ref{37}) the evolution of the horizon
 is given by

 \begin{equation}\label{42}
 \frac{dR_h(t)}{dt}=4 \pi R_h^2 T_{\mu \nu}l^{\mu}l^{\nu}
 \end{equation}

 Hence from (\ref{14}) using (\ref{42}) we can say that validity
 of GSLT does not depend on the type of fluid we are considering-
 the only criteria is whether fluid satisfies the null energy
 condition or not.

\section{Summary and Conclusion:}

In this work we have presented a general formulation of universal
thermodynamics for flat, homogeneous and isotropic FRW model of
the universe. We have started with an interacting $n-$ fluid
system and have shown that the thermodynamics of the universe does
not depend on any individual fluid nor on the interactions, it
depends on the total matter density and thermodynamic pressure of
the fluids i.e. for thermodynamical study we can take the
interacting $n-$fluid system as a single fluid satisfying the
energy conservation relation (\ref{2}). Then we have introduced
the state finder parameters and expressed the restrictions for the
validity of GSLT in terms of these parameters. Also in the
$\{r,s\}-$plane we have shown the feasible regions graphically.\\
In section (III), we have derived the evolution of the horizon for
a general non-static spherically symmetric gravitating system and
we have found (see equation (\ref{38})) that horizon radius
increases or decreases depending on whether null energy condition
is satisfied or not. Then we have calculated the time derivative
of the horizon radius (see equation \ref{42}) for the FRW model.\\
Therefore we can conclude that for a general prescription of
univerasl thermodynamics in FRW model, we need to know the
feasible region in $(r,s)-$ parameter plane and have to examine
whether the resulting single fluid satisfies the NEC or not. For
future work, we shall try, whether similar general features for
universal thermodynamics can be obtained for a general space-time
model of the universe.\\

\end{document}